\documentstyle[aps]{revtex} 

\begin{document} 
  
\centerline{\large\bf On the spectrum of QCD$_{1+1}$ with large numbers of}
\vspace{0.2cm}
\centerline{\large\bf flavours $N_F $ and colours $N_C $ near $N_F /N_C =0$}

\bigskip
\centerline{Michael~Engelhardt\footnote{Supported by DFG 
under contract En 415/1--1.} }
\vspace{0.2 true cm} 
\centerline{\em Institut f\"ur Theoretische Physik, Universit\"at 
T\"ubingen }
\centerline{\em D--72076 T\"ubingen, Germany}
  
\begin{abstract}
QCD$_{1+1}$ in the limit of a large number of flavours $N_F $ and a
large number of colours $N_C $ is examined in the small $N_F /N_C $
regime. Using perturbation theory in $N_F /N_C $, stringent results
for the leading behaviour of the spectrum departing from $N_F /N_C =0$ 
are obtained. These results provide benchmarks in the light of which 
previous truncated treatments of QCD$_{1+1}$ at large $N_F $ and $N_C $ 
are critically reconsidered.
\end{abstract}


\medskip

\section{Introduction}
QCD in one space and one time dimension with large numbers of flavours
$N_F $ and colours $N_C $ has repeatedly attracted attention
\cite{venez}-\cite{cobi}, for various reasons. For one, the success
of the $1/N_C $ expansion provokes the question as to which  of its
aspects are modified if also the number of flavours $N_F $ is taken to
be large; after all, while one may arguably regard $1/N_C $ as a small
number in the real world, this is much harder to justify for $N_F /N_C $.
Compared with the one flavour, many colour case \cite{hoof}, which is
exactly solvable for $N_C \rightarrow \infty $ and yields a Regge
trajectory of mesons consisting strictly of two partons, the many
flavour model is considerably more complex. The suppression of
quark-antiquark pair creation engendered by the large $N_C $ limit
\cite{hoofn} is offset by the increasing number of vacuum polarisation
options as $N_F $ becomes large. In this way, a nontrivial limit of
$N_C \rightarrow \infty $ at constant $N_F /N_C $ emerges which allows
for a much wealthier spectrum than the one-flavour limit. While quark
exchange interactions between different flavour singlet excitations
are still suppressed \cite{self}, pair creation and annihilation effects
within flavour singlets lead to these particles becoming complicated 
mixtures of states of different parton number. They resemble chains whose
links are characterized by the quark and the antiquark at either end of
the link being, alternatingly, either coupled to a colour or a flavour
singlet. In this respect, there is a correspondence to QCD$_{1+1}$ with
one flavour, but adjoint colour quarks; also in the latter model,
extended chain structures form due to each quark possessing two
fundamental colour indices which need to be saturated. This not only
shows up in large $N_C $ studies \cite{dalley}-\cite{hashi}, but also
in the thermodynamical behaviour at $N_C =2$, cf. \cite{adj2}.
The correspondence between the multiflavour, fundamental colour models
and the one-flavour adjoint colour models can in fact be put on a
formally exact footing for $N_F = N_C \rightarrow \infty $, where
the massive spectra in the massless quark case can be shown to
coincide \cite{schwimmer}. 

Not least due to this correspondence, there have been repeated numerical
studies of QCD$_{1+1}$ at large $N_F $ and $N_C $; these include ones using 
quark degrees of freedom \cite{self} as well as more recent and elaborate
ones within the bosonization framework \cite{tritt}. However, all these 
investigations use more or less uncontrolled truncations of the theory. The 
purpose of the present note is to assess the severity of these truncations 
by comparing with results obtained in a particular regime which is
under good numerical control, namely the regime of small $N_F /N_C $.
More specifically, the leading behaviour, proportional to $N_F /N_C $,
departing from the 't Hooft meson spectrum \cite{hoof} at $N_F /N_C =0$ is
obtained using a simple perturbative calculation. Some agreement, but
also significant discrepancies are found as compared
with the aforementioned studies, even the more recent ones. This
indicates that, at this stage, these treatments are not yet as reliable
numerically as one might hope for. The results reported here should provide 
valuable benchmarks for future more elaborate studies of QCD$_{1+1}$ 
in the limit of large $N_F $ and $N_C $.

\section{Fock space and perturbation theory}
The following treatment will concentrate on the case of zero quark
mass matrix, which maximally allows for the pair creation and annihilation
effects highlighted further above. Light-cone coordinates are used and,
moreover, only the sector of overall flavour singlet states will be
considered. In view of colour confinement, physical states will be 
composed out of (colour singlet) 't Hooft mesons \cite{hoof}, which are 
created by the quark bilinear operators
\begin{equation}
M^{\dagger }_{Qnab} = \frac{1}{\sqrt{N_C } } \frac{1}{\sqrt{Q} }
\int_{0}^{Q} dp \, \phi_{n} (p/Q) \sum_{i} q^{\dagger }_{ai} (p)
q_{bi} (p-Q)
\end{equation}
where $Q$ denotes the light-cone momentum of the meson, $n$ its
excitation number, $i$ is the colour index, and $a,b$ the flavour
indices. The wave functions $\phi_{n} $ satisfy 't Hooft's eigenvalue
equation\footnote{Note that the coupling constant $g^2 $ used here and
in \cite{self} differs from the one used in \cite{tritt} and \cite{cobi}
by a factor 2.}
\begin{equation}
-\left( \frac{1}{x} + \frac{1}{1-x} \right) \phi_{n} (x)
-\int_{0}^{1} \frac{dy}{(x-y)^2 } \phi_{n} (y) = \frac{2\pi }{g^2 N_C }
\mu_{n}^{2} \phi_{n} (x) \ , \ \ \ \ x\in [0,1]
\label{hoofeq}
\end{equation}
and form an orthonormal complete set of functions on the interval
$[0,1]$ with the boundary conditions $\phi^{\prime } (0) =
\phi^{\prime } (1) =0$; the eigenvalues $\mu_{n}^{2} $ represent the 
invariant square masses of the mesons. The singularity in the Coulomb
propagator in (\ref{hoofeq}) is defined as usual via the principal
value prescription\footnote{For a dynamical regularization of the
Coulomb propagator via gauge field zero modes, cf. \cite{zphya}.}.
By using the properties of quark operators acting on the vacuum, 
which is the perturbative one in light-cone quantization,
\begin{eqnarray}
q_{ai}^{\dagger } (p) |0\rangle \ \ \ \ &\mbox{for}& \ p<0 \\
q_{ai} (p) |0\rangle \ \ \ \ &\mbox{for}& \ p>0
\end{eqnarray}
one can convince oneself that the flavour singlet one-meson states
\begin{equation}
|2K,n\rangle = \frac{1}{\sqrt{N_F } } \sum_{a} M^{\dagger }_{(2K)naa}
|0\rangle
\end{equation}
are normalized as
\begin{equation}
\langle 2K^{\prime },n^{\prime } | 2K,n\rangle =
\delta (2K-2K^{\prime } ) \delta_{nn^{\prime } }
\end{equation}
Likewise, one can define flavour singlet two-meson states\footnote{Note 
that the normalization of these states differs from the one in 
\cite{self} by a factor $1/\sqrt{2} $. There, this factor was instead 
absorbed into the relative momentum wave function of the two mesons. 
It arises when classifying states according to total and relative momenta 
as opposed to the individual momenta of the mesons.}, where one has two 
options of coupling the flavours:
\begin{eqnarray}
|K+Q/2,m;K-Q/2,n\rangle_{SS} &=& \frac{1}{\sqrt{2} N_F } \sum_{a,b}
M^{\dagger }_{(K+Q/2)maa} M^{\dagger }_{(K-Q/2)nbb} |0\rangle 
\label{sing} \\
|K+Q/2,m;K-Q/2,n\rangle_{NN} &=& \frac{1}{\sqrt{2} N_F } \sum_{a,b}
M^{\dagger }_{(K+Q/2)mab} M^{\dagger }_{(K-Q/2)nba} |0\rangle
\label{trip}
\end{eqnarray}
Either both of the mesons are flavour singlets by themselves (\ref{sing}),
or two flavour non-singlets are coupled to an overall singlet (\ref{trip}).
Furthermore, to avoid double-counting of states, a definite ordering of
the meson momenta in the states will be adopted, namely $Q\geq 0$. This
means that for $m\neq n$, one must distinguish between the states
(\ref{sing}),(\ref{trip}) and the corresponding states with $m$ and $n$
exchanged. The singlet-singlet states (\ref{sing}) completely decouple
from other states of lower or equal parton number in the
large $N_C $ limit \cite{self}, and need not be considered further here.
The states (\ref{trip}) of overall momentum $2K$ are normalized as
\begin{equation}
\langle K^{\prime } +Q^{\prime }/2,m^{\prime } ;
K^{\prime } -Q^{\prime }/2,n^{\prime } |K+Q/2,m;K-Q/2,n\rangle
= \delta (2K-2K^{\prime } ) \delta (Q-Q^{\prime } ) 
\delta_{mm^{\prime } } \delta_{nn^{\prime } }
\end{equation}
where the subscript $NN$ on the states is dropped here and in the 
following. The light-cone Hamiltonian of QCD$_{1+1}$ in the light-cone 
gauge reads
\begin{eqnarray}
H &=& -\frac{g^2 }{4\pi } \frac{N_C^2 -1}{N_C } \sum_{a,i}
\int \frac{dk}{k} : q_{ai}^{\dagger } (k) q_{ai} (k) : \nonumber \\
& & -\frac{g^2 }{8\pi } \sum_{a,b,i,j} \int \frac{dq}{q^2 } dk \,
dk^{\prime } \, \left( : q_{ai}^{\dagger } (k) q_{bi} (k^{\prime } +q)
q_{bj}^{\dagger } (k^{\prime } ) q_{aj} (k-q) : \right. \label{ham} \\
& & \ \ \ \ \ \ \ \ \ \ \ \ \ \ \ \ \ \ \ \ \ \ \ \ \ \
\left. + \frac{1}{N_C } : q_{ai}^{\dagger } (k) q_{ai} (k-q)
q_{bj}^{\dagger } (k^{\prime } ) q_{bj} (k^{\prime } +q) : \right)
\nonumber
\end{eqnarray}
It has been normal-ordered with respect to the perturbative vacuum.
In the 't Hooft limit $N_C \rightarrow \infty $, $N_F /N_C \rightarrow 0$,
the meson states defined further above become eigenstates of the
Hamiltonian,
\begin{eqnarray}
H |2K,n\rangle &=& \frac{\mu_{n}^{2} }{4K} |2K,n\rangle \\
H |K+Q/2,m;K-Q/2,n\rangle &=&
\left(\frac{\mu_{m}^{2} }{2K+Q} + \frac{\mu_{n}^{2} }{2K-Q} \right)
|K+Q/2,m;K-Q/2,n\rangle
\label{thoew}
\end{eqnarray}
When $N_F /N_C $ remains finite as $N_C \rightarrow \infty $, the only
non-vanishing matrix elements involving one-meson states are
\begin{eqnarray}
\langle 2K^{\prime },n^{\prime } | H |2K,n\rangle &=&
\delta (2K-2K^{\prime } ) \delta_{nn^{\prime } } \frac{\mu_{n}^{2} }{4K} 
\label{matel11} \\
\langle 2K^{\prime },n^{\prime } | H |K+Q/2,m;K-Q/2,n\rangle &=&
\sqrt{\frac{N_F }{N_C } } \frac{g^2 N_C }{16\pi K^{3/2} }
(1-(-1)^{m+n+n^{\prime } } ) f_{m n n^{\prime } }
\left( \frac{K+Q/2}{2K} \right) \delta (2K-2K^{\prime } )
\label{coup21}
\end{eqnarray}
with the form factor\footnote{In numerical evaluations of the form factor,
the behaviour around the integration point $(x,y) = (v,0)$, at which the 
integrand superficially becomes singular, can be cast into a manifestly 
nonsingular form by using radial coordinates around this point.}
\begin{equation}
f_{m n n^{\prime } } (v) = \frac{1}{\sqrt{v(1-v)} } \int_{0}^{v} dx
\int_{0}^{1-v} dy \, \phi_{m} (x/v) \phi_{n} (y/(1-v) )
\frac{\phi_{n^{\prime } } (x) - \phi_{n^{\prime } } (v+y) }{(v+y-x)^2 }
\label{formfac}
\end{equation}
With these expressions, it is now straightforward to derive the
leading perturbative corrections, of order $N_F /N_C $, to the masses
of the 't Hooft mesons. The perturbed eigenvalues $E(2K,n)$ of the 
Hamiltonian are given by
\[ E(2K,n) \langle 2K^{\prime },n | 2K,n\rangle = \hfill \]
\begin{eqnarray}
& & \frac{\mu_{n}^{2} }{4K} \langle 2K^{\prime },n | 2K,n\rangle 
\label{pertfull} \\
& & + \int_{0}^{\infty } d(2\bar{K} ) \, \int_{0}^{2\bar{K} } dQ \,
\sum_{\bar{m},\bar{n}=0 }^{\infty }
\frac{ \langle 2K^{\prime },n | 
H | \bar{K} +Q/2,\bar{m} ;\bar{K} -Q/2,\bar{n} \rangle
\langle \bar{K} +Q/2,\bar{m} ;\bar{K} -Q/2,\bar{n} | H | 2K,n \rangle }{
\mu_{n}^{2} /4K \, - \, \mu_{\bar{m} }^{2} / (2\bar{K} +Q) \, - \,
\mu_{\bar{n} }^{2} / (2\bar{K} -Q) }
\nonumber
\end{eqnarray}
cf. (\ref{thoew}). This expression deserves some comment for meson 
excitation number $n=2$ and higher, since these states already 
lie above production thresholds for two-meson states containing
massive mesons of lower $n$ (the zeroth 't Hooft meson is massless).
Therefore, degenerate perturbation theory is called for. In this respect,
the cases $n=2$ through $n=4$ are distinct from $n\geq 5$:

The 't Hooft mesons of excitation number $n=2$ through $n=4$ lie in
a continuum of two-meson states with the special property that one
of the mesons in these states always carries excitation number zero,
i.e. either $\bar{m} =0$ or $\bar{n} =0$ in (\ref{pertfull}). Now, the
form factor (\ref{formfac}) satisfies the relations
\begin{equation}
f_{m0n} \left( \frac{\mu_{m}^{2} }{\mu_{n}^{2} } \right) =
f_{0mn} \left( 1 - \frac{\mu_{m}^{2} }{\mu_{n}^{2} } \right) = 0
\end{equation}
shown in the Appendix. Because of this, zeroes of the energy denominator
in (\ref{pertfull}) are cancelled by zeroes of the coupling
matrix elements; i.e., the perturbed Hamiltonian is already diagonal
in the subspace of states (quasi-)degenerate with the 't Hooft mesons
in question. Therefore, these mesons represent legitimate unperturbed states 
of degenerate perturbation theory and (\ref{pertfull}) remains consistent.

By contrast, mesons of excitation number $n\geq 5$ are degenerate with
two-meson states in which both mesons carry nonzero excitation number.
In this case, (\ref{pertfull}) develops poles (this was verified
numerically), and becomes inconsistent as it stands. Instead, the 
single-meson states in question must be mixed with (quasi-)degenerate
higher parton number states already at the unperturbed level, such as to
rediagonalize the Hamiltonian in this subspace. This more complex case 
will not be considered further here; (\ref{pertfull}) was only evaluated
numerically for the 't Hooft mesons $n=1$ through $n=4$, which in view
of the above discussion represent legitimate unperturbed eigenstates even
in the presence of degeneracies with the two-meson continuum.

Inserting (\ref{coup21}) into (\ref{pertfull}), one can cast the 
invariant square masses in terms of the dimensionless quantities
$e_0 (n)$ and $e_1 (n)$,
\begin{equation}
4K \frac{2\pi}{g^2 N_C } E(2K,n) = e_0 (n) + \frac{N_F }{N_C } e_1 (n)
\end{equation}
with
\begin{eqnarray}
e_0 (n) &=& \frac{2\pi}{g^2 N_C } \mu_{n}^{2} \\
e_1 (n) &=& \frac{1}{2} \sum_{\bar{m},\bar{n}=0 }^{\infty } 
\int_{0}^{1} dx \, \frac{ (1-(-1)^{\bar{m} +\bar{n} +n} )^2 }{e_0 (n)
-2e_0 (\bar{m} )/(1+x) - 2e_0 (\bar{n} )/(1-x) }
\left( f_{\bar{m} \bar{n} n} \left( \frac{1+x}{2} \right) \right)^{2}
\end{eqnarray}
The following table summarizes the results of the numerical evaluation
of $e_0 (n)$ and $e_1 (n)$ for the 't Hooft mesons with excitation number 
$n=1$ through $n=4$.
\[
\begin{array}{|c||c|c|c|c|}
\hline
n & 1 & 2 & 3 & 4 \\
\hline\hline
e_0 (n) & 5.88 & 14.14 & 23.08 & 32.30 \\ \hline
e_1 (n) & 5.1 & 12.0 & -30.5 & 9.1 \\ \hline
\end{array}
\]
In the evaluation, the sums over $\bar{m} , \bar{n} $ were truncated at 
$\bar{m} =10 $, $\bar{n} =10 $; at this level of approximation, neglected 
terms are suppressed compared with the sums obtained by a factor $10^{-4} $
(in the case of $n=4$, by a factor $4\cdot 10^{-4} $). Furthermore,
since all neglected terms are negative, the values given for $e_1 (n)$
represent rigorous upper bounds for the exact values.

The results given in in the above table can be compared to results of 
previous truncated numerical studies of the model. Starting with the first
massive state, the studies \cite{tritt} and \cite{cobi} agree on a slope
$e_1 (1) = 5$ as $N_F /N_C $ is increased from zero, which is confirmed by 
the result $e_1 (1) = 5.1$ obtained here. In \cite{self},
this state was discarded as uninteresting for the (not very compelling) 
reason that it exhibited a positive expectation value of the potential 
energy.

Considering further excited states, however, the spectrum displayed 
in \cite{tritt} suggests that masses systematically rise as $N_F /N_C $ 
is increased from zero\footnote{Note that \cite{tritt} gives results only
for the sector of states which become 't Hooft mesons of odd excitation 
number in the limit $N_F /N_C = 0$.}; indeed, it has been speculated 
\cite{tritt} that $N_F /N_C $ essentially acts as a mass in the model. 
This is in qualitative disagreement with the result obtained above, 
where in particular the trajectory associated with the third 't Hooft meson 
decreases very strongly\footnote{In this respect, the results of 
\cite{self} display at least qualitatively correct behaviour.}.
This disagreement is not entirely surprising in view of the delicate
cancellation observed in connection with (\ref{pertfull}) for mesons
with excitation number $n=2$ and higher, which are embedded in a
continuous spectrum of two-meson states. Already slight truncations
in the numerical treatment will destroy this cancellation and will thus
lead to a remixing of 't Hooft's mesons with the two-meson continuum
already at the unperturbed level, such that results at the next order,
i.e. $N_F /N_C $, become completely unreliable. This is certainly what
happened in \cite{self}, and in view of the strong discrepancy between
the result arrived at in the present work and the behaviour displayed 
in \cite{tritt}, it evidently also takes place there. Indeed, the state 
$n=3$ already is claimed to be a strong mixture  of different parton 
number states in \cite{tritt}.

On a more speculative level, the systematic rise of the spectral
trajectories in the bosonization treatments \cite{tritt},\cite{cobi}
may also be tied to the behaviour found in these studies at large 
$N_F /N_C $, where only invariant square masses of the order of $g^2 N_F $ 
are detected. It is tempting to conjecture that a rich spectrum at the lower 
scale $g^2 N_C $ is thus entirely discarded; such a spectrum
would seem to arise naturally in a perturbative treatment similar in 
spirit to the one presented above\footnote{A first, albeit strongly 
truncated, glimpse of a spectrum at the scale $g^2 N_C $ was seen in 
\cite{self}.}. At large $N_F /N_C $, it is initially indeed natural to 
measure energies in units of $g^2 N_F $; consider e.g. recasting 
(\ref{matel11}) and (\ref{coup21}) in units of $g^2 N_F $ instead of 
$g^2 N_C $. Then, formally, (\ref{matel11}) is of the order $N_C /N_F $ 
and (\ref{coup21}) is of the order $\sqrt{N_C /N_F } $, whereas the 
coupling of two-meson states to two-meson states (given explicitly e.g. 
in \cite{self}) is of order one\footnote{Note that no higher orders arise 
because acting with the Hamiltonian on a multi-meson state only modifies 
up to two mesons in that state.}. Thus, in the large $N_F /N_C $ limit, one 
would regard the latter as the unperturbed Hamiltonian and treat 
(\ref{matel11}) and (\ref{coup21}) as perturbations in $N_C /N_F $. 
Therefore, the unperturbed problem initially yields a spectrum of energies 
proportional to $g^2 N_F $, as indeed happens in the bosonization approach
\cite{cobi95}-\cite{cobi}. However, this is only half the story;
it should be realized that in this scheme there is a rich spectrum of
(unperturbed) massless states. Among them are e.g. all of 't Hooft's
mesons. Now, even if one is only interested in the massive spectrum
of the model, one should not prematurely discard these zero energy
states\footnote{Note that the treatment of the large $N_F $ limit in 
\cite{cotri} invokes a saddle-point argument based on the magnitude of 
$N_F $; great care must be exercised in justifying such an argument in the 
presence of (quasi-) zero modes.}, since their 
masses presumably are not protected against corrections once one includes 
the perturbations in $N_C /N_F $. Already at the next order, i.e. 
$g^2 N_F \cdot N_C /N_F $, single meson states acquire corrections
from two sources: First order perturbation theory in (\ref{matel11}),
yielding precisely 't Hooft's masses, and second order perturbation theory
in (\ref{coup21}), coupling to the two-meson states\footnote{Of course,
additional care must be taken if massless unperturbed two-meson
states occur in this scheme.}. Thus, in the limit 
$N_F /N_C \rightarrow \infty $, there is a sound basis for the conjecture 
that a well-defined spectrum of invariant square masses of the order of 
$g^2 N_C $ emerges.

In summary, it seems that more work is needed before a coherent picture 
of QCD$_{1+1}$ with a large number of colours and flavours can be presented; 
the truncation schemes hitherto applied leave room for error even on a 
qualitative level. The results presented here for the first time
provide stringent numerical control over the behaviour of the spectrum
departing from the 't Hooft limit $N_F /N_C = 0$, albeit in a very limited
range of the parameter $N_F /N_C $, and only for the low-lying 't~Hooft 
mesons of excitation number $n=1$ through $n=4$. These results may provide 
useful benchmarks for future improved studies of the model. 

\section*{Appendix}
Consider the form factor $f_{m0n} (v)$, cf. eq. (\ref{formfac}). 
Inserting the special form of the zeroth 't Hooft meson wave function, 
$\phi_{0} (x) = 1$, this form factor satisfies
\begin{eqnarray}
\sqrt{v(1-v)} f_{m0n} (v) &=&
\int_{0}^{v} dx \int_{0}^{1-v} dy \, \phi_{m} (x/v)
\frac{\phi_{n} (x) - \phi_{n} (v+y)}{(v+y-x)^2 } \\
&=& \int_{0}^{v} dx \, \phi_{m} (x/v) \phi_{n} (x) \int_{0}^{1-v} dy \,
\frac{1}{(v+y-x)^2 } \\
& & - v \int_{0}^{1} dz \int_{-v}^{1-v} dy \, 
\frac{\phi_{m} (z) \phi_{n} (v+y)}{(v+y-vz)^2 }
+ v \int_{0}^{1} dz \int_{-v}^{0} dy \,
\frac{\phi_{m} (z) \phi_{n} (v+y)}{(v+y-vz)^2 } \\
&=& \int_{0}^{v} dx \, \phi_{m} (x/v) \phi_{n} (x)
\left( \frac{1}{v-x} -\frac{1}{1-x} \right) \\
& & - v \int_{0}^{1} dz \int_{0}^{1} dy \,
\frac{\phi_{m} (z) \phi_{n} (y)}{(y-vz)^2 }
+ \frac{1}{v} \int_{0}^{1} dz \int_{0}^{v} dy \,
\frac{\phi_{m} (z) \phi_{n} (y)}{(y/v-z)^2 }
\label{refap}
\end{eqnarray}
Using 't Hooft's equation (\ref{hoofeq}) to carry out the $y$-integral
in the first term in (\ref{refap}) and, likewise, the $z$-integral
in the second term in (\ref{refap}), one arrives at
\begin{eqnarray}
\sqrt{v(1-v)} f_{m0n} (v) &=&
v \int_{0}^{1} dz \, \phi_{m} (z) \phi_{n} (vz)
\left( \frac{1}{v-vz} -\frac{1}{1-vz} \right) \\
& & -v \int_{0}^{1} dz \, \phi_{m} (z) 
\left( -\frac{1}{vz} -\frac{1}{1-vz} - \frac{2\pi }{g^2 N_C } \mu_{n}^{2}
\right) \phi_{n} (vz) \\
& & +\frac{1}{v} \int_{0}^{v} dy \, \phi_{n} (y)
\left( -\frac{v}{y} -\frac{1}{1-y/v} - \frac{2\pi }{g^2 N_C } \mu_{m}^{2}
\right) \phi_{m} (y/v) \\
&=& \int_{0}^{1} dz \, \phi_{m} (z) \phi_{n} (vz)
\left( \frac{1}{1-z} - \frac{v}{1-vz} +\frac{1}{z} +\frac{v}{1-vz}
+\frac{2\pi }{g^2 N_C } v \mu_{n}^{2} \right) \\
& & +\int_{0}^{1} dz \, \phi_{n} (vz) \phi_{m} (z) 
\left( -\frac{1}{z} -\frac{1}{1-z} - \frac{2\pi }{g^2 N_C } \mu_{m}^{2}
\right) \\
&=& \frac{2\pi }{g^2 N_C } 
\int_{0}^{1} dz \, \phi_{m} (z) \phi_{n} (vz)
(v\mu_{n}^{2} - \mu_{m}^{2} )
\end{eqnarray}
At the particular value $v=\mu_{m}^{2} / \mu_{n}^{2} $, the form factor
$f_{m0n} (v)$ thus exhibits a zero. The case of 
$f_{0mn} (1-\mu_{m}^{2} / \mu_{n}^{2} )$ can be treated in complete analogy.

\end{document}